\documentclass[prd,nofootinbib,superscriptaddress,twocolumn]{revtex4}

\usepackage[mathscr]{eucal}
\usepackage[latin1]{inputenc}
\usepackage{amsmath}
\usepackage{amsfonts}
\usepackage{amssymb}
\usepackage{pictex}
\usepackage{graphicx}

\def\b{{\tt b}}
\def\be{\nopagebreak[3]\begin{equation}}
\def\ee{\end{equation}}
\def\ba{\nopagebreak[3]\begin{eqnarray}}
\def\ea{\end{eqnarray}}

\def\d{{\rm d}}
\def\w{\omega}

\def\H{{\cal H}}

\newcommand{\teta}{\rlap{\lower2ex\hbox{$\,\tilde{}$}}\eta{}}

\def\H{{\cal H}}

\def\Tr{{\rm Tr\,}}

\def\lp{{\ell}_{\rm Pl}}

\def\w{{}^o\!\omega}

\newcommand{\f}{\frac}

\usepackage{enumerate}

\usepackage{colordvi}
\newcounter{mnotecount}[section]

\newcommand{\mnote}[1]

\newcommand{\comment}[1]{}

\def\f{\frac}

\def\epsilon{\varepsilon}

\def\sindb{\sin \left(\f{\delta_b b}{2}\right)}
\def\sintdb{\sin^2\left(\f{\delta_b b}{2}\right)}
\def\cosdb{\cos \left(\f{\delta_b b}{2}\right)}
\def\costdb{\cos^2\left(\f{\delta_b b}{2}\right)} 
\def\sindc{\sin \left(\f{\delta_c c}{2}\right)}

\def\cosdc{\cos \left(\f{\delta_c c}{2}\right)}

\def\db{\delta_b}
\def\dc{\delta_c}

\begin{document}
\preprint{\vbox{\baselineskip=12pt \rightline{IGC-08/2-?}
}}

\title{Loop quantization of the Schwarzschild interior revisited
}

\author{Alejandro Corichi}\email{corichi@matmor.unam.mx}
\affiliation{Centro de Ciencias Matem\'aticas,
Universidad Nacional Aut\'onoma de
M\'exico, UNAM-Campus Morelia, A. Postal 61-3, Morelia, Michoac\'an 58090,
Mexico}
\affiliation{Center for Fundamental Theory, Institute for Gravitation and the Cosmos,
Pennsylvania State University, University Park
PA 16802, USA}
\author{Parampreet Singh}
\email{psingh@phys.lsu.edu}
\affiliation{Department of Physics and Astronomy, Louisiana State University, Baton Rouge, 
LA 70803, USA}

\begin{abstract}
The loop quantization of the Schwarzschild interior region, as described by a homogeneous
anisotropic Kantowski-Sachs model, is re-examined. 
As several studies of different --inequivalent-- loop quantizations have shown,
to date there exists no fully satisfactory quantum theory for this model. This fact 
poses challenges to the validity of some scenarios to address the black hole information problem. 
Here we put forward a novel viewpoint to construct the quantum theory that 
builds from some of the models available in the literature. 
The final picture is a quantum theory that is both independent of any auxiliary structure and possesses a 
correct low curvature limit. It represents a subtle but non-trivial modification of the 
original prescription given by Ashtekar and Bojowald. It is shown that the quantum gravitational 
constraint is
well defined past the singularity and that its effective dynamics possesses a bounce into
an expanding regime. The classical singularity is avoided, and a semiclassical spacetime
satisfying vacuum Einstein's equations is recovered on the ``other side" of the bounce.
We argue that such metric represents the interior region of a white-hole spacetime, but for which
the corresponding ``white-hole mass" differs from the original black hole mass. Furthermore, we find that 
the value of the white-hole mass is proportional to the third power of the starting black hole mass. 
\end{abstract}

\maketitle

\section{Introduction}

The singularity theorems of general relativity tell us that, under generic conditions,
spacetime singularities {\it will} form. 
From the study of solutions to Einstein's equations we have learned that spacelike 
singularities arise in two physical situations: in cosmological scenarios where we associate them with either the Big Bang or the Big Crunch, and in the interior regions of event horizons, that is, inside black holes. 
Just as in the cosmological scenario, FLRW cosmological spacetimes are taken as paradigmatic examples of space-times possessing a singularity, the
Schwarzschild spacetime is the standard example of those singularities of the second kind.
The standard interpretation within general relativity,
when such singularities appear, is that the
description one is using of the physical situation breaks down; general relativity is no longer valid. In terms of physical quantities, such as geometric scalars, these singularities are manifest when some of such quantities grow un-boundedly. One can therefore expect that in these physical situations, quantum gravitational effects will take over and become dominant. The problem is that we do not yet possess a complete theory that describes such quantum phenomena.
A very conservative approach is to apply `standard' quantization techniques to highly
symmetric classical configurations, such as homogeneous spacetimes in the cosmological context and for the interior of black holes. This is the strategy that we shall consider in this manuscript. Since the interior (black hole) Schwarzschild solution
can be seen as a particular example of a contracting anisotropic Kantowski-Sachs model, one natural question is to consider its quantization with the aim of probing what the fate of the
final classical singularity may be.

In recent years, new quantization techniques have been applied to these minisuperpace
scenarios. These techniques, motivated by loop quantum gravity (LQG) \cite{lqg}, are based on
non-regular representations of the canonical commutation relations that make these quantum systems inequivalent to the standard Schr\"odinger representation,  already at the kinematical level. In the cosmological scenario the resulting formalism is known as loop quantum cosmology (LQC) \cite{lqc}. The main features that distinguishes these quantum minisuperspace models from the standard Wheeler-De Witt (WDW) models is that, for simple systems, the singularity is generically avoided and the quantum evolution continues past the `would be singularity' in a unitary way \cite{aps2,slqc}. These contrasting features have been rigorously understood via various analytical \cite{lqc} and numerical investigations in LQC \cite{ps12}. It is therefore natural to ask whether similar results emerge when these technique are applied to the Schwarzschild interior. The pioneer work that embarked on this task was
Ref.~\cite{aa:mb}, and several other works followed \cite{Modesto,b-k,Campiglia,jp}. 

Unlike the regular representations, where the Stone-Von Neumann theorem warranties uniqueness
(up to unitary equivalence), within the realm of loop quantizations (sometimes referred also as `polymer quantizations'), there are infinite inequivalent representations of the quantum constraints. In simple cases, this freedom
translates into the liberty of choosing the basic variables for the quantization.
In the original treatment of the black hole interior \cite{aa:mb}, the strategy followed  a particular quantization of isotropic FLRW models \cite{abl,aps1}. As in the 
early quantization of cosmological models in LQC, the quantizations of black hole spacetime given in Refs. \cite{aa:mb, Modesto, Campiglia, jp} suffer from lack of 
independence of the the auxiliary structure, such as the `size' of the fiducial cell, needed for the Hamiltonian description. This results in spurious quantum gravitational effect at ultra-violet and infra-red scales \cite{cs2}. Further, in these quantizations singularity resolution can not be consistently identified with a curvature scale \cite{cs3,ksbound}. 
A new proposal motivated by the `improved quantization' in LQC \cite{aps2} was put forward by B\"ohmer and Vandersloot \cite{b-k} (see also Ref.\cite{mb_cartin_khanna}).
Though this quantization is free from the auxiliary structure and results in universally bounded expansion and shear scalars \cite{ksbound}, it has the following limitation. 
The quantization leads to `quantum gravitational effects' at the horizon due to the coordinate singularity. Nevertheless, B\"ohmer and Vandersloot's prescription turns out to be 
free from these effects when the quantization of Kantowski-Sachs spacetime is performed in the presence of matter. Interestingly, after the would-be singularity is avoided the spacetime 
corresponds to a `charged' Nariai spacetime in B\"ohmer and Vandersloot's prescription \cite{charged,ksphen}, a conclusion which remains unchanged for the vacuum case and hence for the Schwarzschild interior.   

Given that the quantizations of black hole spacetimes based on the original LQC suffer from auxiliary structures dependence and the lack of a consistent singularity resolution scale, and the attempt based on `improved quantization' in LQC results in large `quantum gravity effects' at low curvatures near the horizon, none of the available quantizations of the Schwarzschild interior in LQG can be deemed satisfactory.
The purpose of this manuscript is to advance a new viewpoint 
to construct a quantization that is free from these limitations. The main new ingredient in the construction is the realization that, in order to define the Hamiltonian
formulation, it is necessary to introduce a physical length scale $r_o$ into the problem. This
is a true scale that, in the classical limit, selects a unique classical solution and becomes nothing but 
the Schwarzschild radius. This scale is to be contrasted with the auxiliary length $L_o$ that fixes the 
fiducial cell needed for defining the symplectic structure. The latter being an auxiliary structure has 
no physical significance and should be absent in the final quantum theory.

The proposal that we shall describe in detail is motivated by recent considerations regarding the loop quantization of symmetry reduced models \cite{cs2,cs3}. There, it was proposed that rather natural physical criteria be used to select various choices in the quantization procedure. In particular one should require that the final quantum theory be independent of any auxiliary structure, have 
a well defined high curvature regime and recover the classical theory at low curvatures. In the case of the black hole interiors, no such construction exists to date. Here 
we are able to construct a quantum theory that satisfies all these criteria and, even when it resembles the original treatment of \cite{aa:mb}, it possesses subtle and important differences. 
These are manifested in the
inputs necessary for regulating the curvature in the Hamiltonian constraint using holonomies which capture the fundamental discrete nature of the underlying quantum geometry. We notice that different prescriptions need to be specified for the two anisotropic directions, with distinct dependence on the two
scales at hand, namely the (new) physical scale $r_o$ and the auxiliary $L_o$. 
This subtle difference makes the resulting quantum theory 
singularity free and independent of  $L_o$, thus removing dependence on any auxiliary structure. A study of the semiclassical `effective dynamics' also shows  that there are no spurious quantum gravity effects appearing at low curvatures. From this viewpoint, this represents the first fully consistent quantization for black hole interiors using loop quantization methods which has a well defined and consistent ultra-violet regime where the singularity resolution occurs, and which is in agreement with general relativity in the infra-red limit. It should be noted that we arrive at this description without invoking ideas from improved dynamics of LQC which involved a subtle
change of the original discretization variable. 
The classical singularity is avoided, and replaced by a `quantum bounce' to a new branch. The spacetime does not end at the singularity. An interesting feature of the effective spacetime that arises
after the bounce is that it represents the (expanding) interior region of a white hole, but of a different mass than the original black hole. Furthermore, there appears to be a cubic powerlaw relation between the two sets of masses. It should be noted that the difference in the white hole mass and the starting black hole mass is also 
present in the Ashtekar-Bojowald prescription as was first shown in Ref. \cite{b-k}. But, there is a striking difference in physics in comparison to our analysis. Unlike the quantization introduced in this manuscript, the white hole mass in the Ashtekar-Bojowald prescription depends on the fiducial length $L_o$, and not on the starting black hole mass. Thus, while the white hole mass in our prescription is proportional to the cubic power of the initial black hole mass and independent of any fiducial structure,   in the Ashtekar-Bojowald prescription it can take any arbitrary value through its $L_o$ dependence.

The structure of the paper is as follows. In Sec.~\ref{sec:2}, we recall the classical minisuperspace in terms of connections and triads that we are considering and introduce the main assumption underlying the construction of the quantum theory. Sec.~\ref{sec:3} is devoted to the study of the classical solution for the equations and their space-time interpretation. In Sec.~\ref{sec:4} we consider the loop quantization of these models and arrive at a quantum difference equation that selects the physical states of the theory. 
In Sec.~\ref{sec:5} we consider
the effective dynamics one expects to recover from the quantum theory and study its phenomenological 
implications. Here we  discuss the boundedness of expansion and shear scalars in the effective spacetime 
description when the would-be central singularity is approached. We show the independence of the physics 
on the underlying fiducial structures and discuss the relationship between the starting black hole mass 
and the white hole mass.  We conclude the article with a summary of results and a discussion 
in Sec.~\ref{sec:6}.

\section{Preliminaries}
\label{sec:2}

The portion of the 
Schwarzschild spacetime corresponding to the interior region of the black hole
can described by a homogeneous cosmological model given by a  Kantowski-Sachs 
geometry with symmetry group $\mathbb{R} \times SO(3)$. 
This corresponds to a foliation in which the 
3-manifolds with topology $M = \mathbb{R} \times \mathbb{S}^2$ approach
a null-surface, the horizon, on the past and a space-like singularity on the
future. As in the previous cases studied within minisuperspace Hamiltonian formulation, 
it is a standard procedure to
define a fiducial metric ${}^o\!q_{ab}$ on the
3-manifold $M = \mathbb{R} \times \mathbb{S}^2$ and define the
corresponding compatible triad ${}^o\!e^i_a$ and co-triad ${}^o\!\omega^a_i$. 
However, unlike the isotropic and homogeneous FLRW and Bianchi-I models, there is only one 
non-compact direction and thus one has to introduce a line interval $L_o$
as an auxiliary structure to define the phase space.
The choice of fiducial metric that we will here make, even when rather similar to
that of \cite{aa:mb} (and used in all subsequent articles  
\cite{Modesto,b-k,mb_cartin_khanna,Campiglia}) is, however, motivated by different 
considerations. In \cite{aa:mb}, the fiducial metric on the sphere was chosen to be a
unit sphere. Even when this choice is mathematically consistent, one needs to specify 
a scale, in this case with dimensions of area that, together with the dimension-full 
coordinate $x$, yield a fiducial metric with consistent dimensions. The viewpoint that
we shall adopt here is that, in order to perform a Hamiltonian analysis in the absence 
of a natural scale in GR, one has to specify it externally \emph{as a boundary condition}. 
This situation is not new in black hole physics. For instance,
as part of the definition of an \emph{isolated horizon} \cite{IH}, one has 
to specify the precise values of all the multipole moments of the horizon. For the simplest 
case of spherical horizons (of Type I in the standard terminology), this means fixing the
value of the area $a_o$ of the horizon. The classical Hamiltonian theory is thus dependent 
on this external parameter, and this is carried over to the quantum regime. 

Our concrete proposal is  to extend this viewpoint to the formalism that we are here considering,
and fix a classical scale by asking that the area $a_o$ of $\mathbb{S}^2$ be equal 
to $a_o:=4\pi\,r^2_o$. The fiducial metric takes the form:
\be
\d s_o^2:=\d x^2 + r^2_o(\d \theta^2 +\sin^2\theta\,\d \phi^2)
\ee
with determinant ${}^oq = r^4_o\,\sin^2\theta$. This externally prescribed length scale will 
turn out to be the Schwarzschild radius, thus providing a physical parameter that defines
the configurations under consideration. Namely, just as in the analysis of entropy for isolated horizons
 where one chooses a given fixed area from the start, the system under consideration here will be the 
 dynamics of a Kantowski-Sachs 
cosmology with the interpretation of being the interior region of a black hole of a given area $a_o$.
In practical terms, this does not represent a limitation, since at the end of the day one can let $a_o$
become a free parameter, so we have a description of all possible black hole interiors. We have been
careful in explaining the role of $r_o$ since it represents the main conceptual difference from other treatments, and the one that will allow for a consistent formulation.

In order to define the symplectic structure,  one has to 
specify a finite interval for the coordinate $x$ in $\mathbb{R}$.  
The usual procedure is to restrict the
range of $x$ to the interval $[0,L_o]$. With this choice the fiducial volume of the
fiducial cylinder is equal to $V_o=4\pi\,r^2_o\,L_o$. With our choices, this quantity has 
the `right dimensions' of volume.

Utilizing the symmetries of the spacetime (and after imposing the Gauss constraint), the 
connection and the triad can be written as follows \cite{aa:mb}:
\be
A^i_a \, \tau_i \, \d x^a \, = \, \bar c \, \tau_3 \, \d x + \bar b\,r_o \, \tau_2 \d \theta 
- \bar b \,r_o\, \tau_1 \sin \theta \, \d \phi + \tau_3 \cos \theta \, \d \phi
\ee
and
\be
E^a_i \, \tau^i \f{\partial}{\partial x^a} \, =  \, \bar p_c \,r_o^2\, \tau_3 \, \sin \theta 
\, \f{\partial}{\partial x} + \bar p_b\, r_o \, \tau_2 \, \sin \theta \, 
\f{\partial}{\partial \theta} - \bar p_b\, r_o \, \tau_1 \,  \f{\partial}{\partial \phi} ~
\ee
with fiducial triad and cotriad:
\be
{}^o\!A^i_a \, \tau_i \, \d x^a \, = \, \tau_3 \, \d x + r_o \, \tau_2 \d \theta - r_o\, 
\tau_1 \sin \theta \, \d \phi + \tau_3 \cos \theta \, \d \phi
\ee
and
\be
{}^oE^a_i \, \tau^i \f{\partial}{\partial x^a} \, =   r_o^2\, \tau_3 \, \sin \theta \, 
\f{\partial}{\partial x} +  r_o \, \tau_2 \, \sin \theta \, \f{\partial}{\partial \theta} -
 r_o \, \tau_1 \,  \f{\partial}{\partial \phi} ~.
\ee

The metric for the particular case of the Schwarzschild interior spacetime is
\be
\d s^2 = - N^2 \d t^2 + \f{\bar p_b^2}{|\bar p_c|} \, \d x^2 + |\bar p_c|\,r_o^2 \, 
(\d \theta^2 + \sin^2 \theta\, \d \phi^2) ~
\ee
Thus $\bar p_\b$ and $|\bar p_c|$ are related to components of the standard 
Schwarzschild metric as
\be
\f{\bar p_b^2}{|\bar p_c|} = (2 m/t - 1), ~~~~ |\bar p_c|\,r_o^2 = t^2.
\ee
where $m=GM$ and $M$ is the  ADM mass of the spacetime.
The triads $\bar p_b$ and $\bar p_c$ are dimensionless, the latter being equal to unity at the horizon by the choice of parameter $r_o$ (with dimensions of mass parameter): $r_o = 2 m$.
The symplectic structure expressed in terms of the
conjugate pairs $(\tilde b, \tilde p_b)$ and $(\tilde c, \tilde p_c)$ is not invariant 
under the change of fiducial metric. We can introduce
\be\label{cbpair}
c = L_o \, \bar c, ~~ p_c = r_o^2\,\bar p_c, ~~ b = r_o\,\bar b, ~~ p_b = r_o\,L_o \, \bar p_b
\ee
which satisfy
\be\label{pbs}
\{c,p_c\} \, = \,2 G \gamma, ~~~~ \{b,p_b\} \, = \,  G \gamma ~.
\ee
There are two underlying freedoms associated with auxiliary structures of which the theory should
be invariant. The redefinition (\ref{cbpair})  takes care of the freedom to rescale coordinates
(for eg. $x \rightarrow \lambda x$) keeping the metric invariant. However, freedom to 
rescale size of the interval of integration $L_o$ exists, and as in the cosmological model 
one of the primary tasks will be to identify suitable variables for the phase space. 
Under the rescaling of 
$L_o \rightarrow \alpha L_o$:
\ba\label{schw_trans}
c \, \rightarrow \, c' \, = \, \alpha \, c, && \nonumber ~~ p_c \, \rightarrow \, p_c' \, = 
\, p_c, ~~\\
b \, \rightarrow \, b' \, =  \, b, && ~~ p_b \, \rightarrow \, p_b' \, = \, \alpha \, p_b ~.
\ea
Note that in our model, there is no underlying freedom to change $r_o$. As stated before it amounts to considering a different physical situation.  This is quite different from the quantity $L_0$ that is an auxiliary
(and arbitrary) construct with no physical meaning.

\section{Classical Solutions}
\label{sec:3}

The classical Hamiltonian constraint in terms $(b,p_b)$ and $(c,p_c)$ 
can be written as \cite{aa:mb}
\be\label{h_cl}
C_{\mathrm{class}} = - \f{8 \pi N \mathrm{sgn} (p_c)}{\gamma^2}\left((b^2 + \gamma^2) \f{p_b}{\sqrt{|p_c|}} + 2 b c |p_c|^{1/2} \right) ~.
\ee
It is related to the classical Hamiltonian as: $H_{\mathrm{class}} = C_{\mathrm{class}}/16 \pi G$. Choosing $N = \gamma \mathrm{sgn} (p_c) |p_c|^{1/2}/b$ in $H_{\mathrm{class}}$ and using (\ref{pbs}), we can find the equations of motion for the 
phase space variables:
\be
\dot{b}=-\frac{1}{2}\,\left(b+\f{\gamma^2}{b}\right)\, , ~~ \dot{p}_b = \frac{p_b}{2}\,\left(1-\f{\gamma^2}{b^2}\right)\, 
\ee
and 
\be
\dot{c}= - 2c\, , ~~ \dot{p}_c = 2 p_c ~
\ee
where the derivative is with respect to `time' $T$. 
Integrating the equations for $b$, $p_b$ and $p_c$ and using the constraint equation
\be
c = - \frac{p_b}{2 p_c}\left(b+\f{\gamma^2}{b}\right)\, .
\ee
we obtain 
\ba\label{class_sol_T}
p_b(T)&=&p_b^{(o)}\, e^T\,\sqrt{e^{-(T-T_o)} - 1}\, ,\\
p_c(T)&=&p_c^{(o)}\,e^{2T}\, ,\\
b(T)&=&\pm \gamma\,\sqrt{e^{-(T-T_o)}-1}\, \\
{\mathrm{and}} ~~~~~ c(T) &= & c_o\,e^{-2T} ~.
\ea
It is convenient to make a change of variables from dimensionless $T$ to dimensionfull 
$t:=l_o\,e^T$, with $l_o$ as a length scale to be determined. We also identify the free parameter 
in $b$ and $p_b$, denoted $T_o$ as
 $\tilde m=(l_o/2)\, e^{T_o}$.  With this change of variables, the solutions take the form:
\ba
b(t) &=& \pm \gamma\,\sqrt{(2 \tilde m-t)/t}~,\\
p_b(t) &=& \frac{p^{(o)}_b}{l_o}\,\sqrt{t(2 \tilde m-t)}\, \\
p_c(t)&=&\pm\frac{p_c^{(o)}}{l_o^2}\,t^2\, \\
{\mathrm{and}} ~~~~~ c(t) &=& \mp \gamma \tilde m\frac{p_b^{(o)}}{p_c^{(o)}}\,\frac{l_o}{t^2}\, ~.
\ea

So far, we have four free parameters, two of which have to be fixed to end up 
with only two
(corresponding to the true degrees of freedom in the canonical formulation). The first simplification pertains to the 
freedom we have in choosing the initial time $T_o$ (given that the constraint generates
constant translations in $T$, and one has to reduce this freedom). Thus one can chose
without losing generality $T_o=0$ which implies $l_o=2 \tilde m$. Since the Schwarzschild radius, $r_o$ is
a natural physical length scale in the model we further identify 
$l_o = r_o$ which implies $\tilde m = m$, the mass parameter of the black hole.
Note that $|p_c|$ has the interpretation, in terms of the 
spacetime metric of the geometric radius of the homogeneous spheres. Thus, we would like 
to associate the parameter $t$ with this quantity. This implies the choice 
$p_c^{(o)}:=r_o^2=4m^2$. With these identifications the solutions can be 
rewritten as
\ba
b(t) &=& \pm \gamma\,\sqrt{(2 m-t)/t}~,\\
p_b(t) &=& \frac{p^{(o)}_b}{2 m}\,\sqrt{t(2 m-t)} \,, \, \\
p_c(t)&=&\, \pm t^2\, \\
{\mathrm{and}} ~~~~ c(t) &=& \mp \f{\gamma}{2} \frac{p_b^{(o)}}{t^2}\, ~.
\ea

We are thus left with only two free quantities, namely $(m,p_b^{(o)})$. 
We further note that 
the spacetime metric only knows of the one independent parameter ($m$). This can be 
seen by considering the $g_{xx}$ component that is given by,
\ba
g_{xx} & = &  \frac{p^2_b(t)}{|p_c(t)|L^2_o} = \frac{(p_b^{(o)})^2}{|p_c^{(o)}|L_o^2}\,
\left(\frac{2m}{t}-1\right)\nonumber \\
& = & \frac{(p_b^{(o)})^2}{4m^2\,L_o^2}\,\left(\frac{2m}{t}-1\right)
\, ,
\ea
From which one can fix $p_b^{(o)}$ to be $p_b^{(o)}=2m\,L_o$, in order to have the standard 
form of the line element. Note that for any other choice of $p_b^{(o)}$, say 
$\tilde{p}_b^{(o)}= \lambda^2\, p_b^{(o)}$
one could still 
bring the metric to the standard form by a constant rescaling of the coordinate 
$x$ by defining $x'=\lambda x$ \footnote{However such a rescaling, even when valid
from the spacetime perspective, is not a canonical transformation from the Hamiltonian
perspective, given that $L_o$ is also rescaled by $\lambda$ when changing coordinates. This illustrates the fundamental difference between the Hamiltonian and the spacetime approaches to this system.
This discussion corrects some statements of \cite{aa:mb} and \cite{b-k} 
that were based on a slightly different parametrizations.}.

There is another  way of solving the equations of motion that is illustrative to 
understand the behavior. First note that there is a natural constant of the motion
given by $h=2c\,p_c$ (in the standard terminology it is a Dirac observable). With this 
identification we see that the Hamiltonian constraint can be rewritten:
\be
h=p_b\,\left(b + \frac{\gamma^2}{b}\right)\, ,
\ee
Thus, we can interpret $h$ as a reduced Hamiltonian that dictates the motion for the 
$(b,p_b)$ part of the phase space, that is decoupled from the $(c,p_c)$ sector.
It is also a first integral that gives us the relational dynamics of $p_b$ as function of
$b$:
\be
p_b=\frac{h\,b}{b^2+\gamma^2}
\ee
There is only one free parameter in this dynamics, namely the value of $h$.
In the $(c,p_c)$ sector since $c=h/2p_c$, we only have to fix one constant
(for instance $p_c^{(o)}$.). Thus we recover the two degrees of freedom. It is now 
immediate to identify and interpret the dynamical evolution, with parameter $t$ 
(or $|p_c|^{1/2}$), in the phase space.
For $t=2m$, that in the spacetime pictures represents the horizon, we have that
$b=p_b=0$ and $c=\gamma\,L_o/4m$. As $t$ approaches zero, $b$ and $c$ grow unboundedly,
and $p_b$ increases from $b=0$ to $b=\gamma$ where it reaches its maximum value 
$p_b^{\rm max}=r_o\,L_o/2$, and then becomes a monotonically decreasing function 
that approaches zero as $t\to 0$.

\section{Loop Quantization}
\label{sec:4}

The elementary variables for the quantization are the holonomies of the 
connections $c$ (considered over edges labelled by $\tau$ in $x$ direction) and $b$ (considered
over edges labelled by $\mu$  in $\theta$ and $\phi$ directions) 
\be\label{hx}
h_x^{(\tau)} = \cos(\tau c/2) + 2\, \tau_3 \, \sin(\tau c/2) ~,
\ee
\be\label{htheta}
h_\theta^{(\mu)} = \cos(\mu b/2) + 2\, \tau_2 \, \sin(\mu b/2) ~
\ee
and
\be\label{hphi}
 h_\phi^{(\mu)} = \cos(\mu b/2) - 2\, \tau_1 \, \sin(\mu b/2) ~.
\ee 

The holonomies generate an algebra of the almost periodic functions with 
elements of the form $\exp(i(\mu b + \tau c)/2)$  and the resulting kinematical
Hilbert space is a space of square integrable functions on the Bohr compactification of $\mathbb{R}^2$:${\cal H} = L^2(\mathbb{R}_{\mathrm{Bohr}}, \d\mu_B)$. The 
eigenstates of $\hat p_b$ and $\hat p_c$ are:
\be\label{pb_pc_ev}
\hat p_b \, |\mu,\tau\rangle = \f{\gamma \lp^2}{2} \, \mu \, |\mu,\tau\rangle, ~~ \hat p_c \, |\mu,\tau\rangle = \gamma \lp^2 \, \tau \, |\mu,\tau\rangle ~.
\ee
which satisfy $\langle \mu', \tau'|\mu, \tau\rangle = \delta_{\mu \mu'} \delta_{\tau \tau'}$.

In order to quantize the gravitational constraint we treat extrinsic curvature
$K^i_a = \gamma^{-1} (A^i_a - \Gamma^i_a)$ as the connection and consider its curvature ${}^o\!F_{ab}^k$, using which
the classical constraint can be written as
\be
C_{\mathrm{Ham}} = - \int \d^3 x \, e^{-1} \varepsilon_{ijk} E^{ai} E^{bj} (\gamma^{-2} \, {}^o\!F_{ab}^k - \Omega_{ab}^k) ~.
\ee
Here $\Omega = -\sin \theta \tau_3 \d \theta \wedge \d \phi$ is the curvature corresponding to the spin connection 
$\Gamma = \cos \theta \, \d \phi$. At the equator
of $\mathbb{S}^2$, the spin connection vanishes and $K^i_a = \gamma^{-1} A^i_a$.
Holonomies of extrinsic curvature along $x$, $\theta$ and $\phi$ directions then turn out to be equal to the holonomies of connection (\ref{hx},\ref{htheta},\ref{hphi}) when computed from equator.
To be precise, we consider loops in $x-\theta$, $x-\phi$ an $\theta-\phi$ planes. The edge along $x$ direction in $\mathbb{R}$ has length $\delta_c \ell_c$ with $\ell_c = L_o$ and the edges along longitudes and equator of $\mathbb{S}^2$
each having length $\delta_b \ell_b$ with $\ell_b = r_o$.

The term proportional to inverse triad can be casted in terms of holonomies by
using an identity on the classical phase space:
\ba\label{term1}
\varepsilon_{i j k} e^{-1} E^{a j} E^{b k} &=& \nonumber  \sum_k \f{{}^o\!\varepsilon^{a b c} \, 
{}^o\omega_c^k}{2 \pi \gamma G \delta_{(k)} \ell_{(k)}}\\
&& \hskip-0.4cm\times  \, \mathrm{Tr} \left(h_k^{(\delta_{(k)})} \, \{(h_k^{(\delta_{(k)})})^{-1},V\} \tau_i\right)
\ea
where $\delta_{(i)}$ (and $\ell_{(i)}$) correspond to $\delta_b$ or $\delta_c$ (and similarly $L_o$ or $r_o$) depending on the edge over which a holonomy is computed. Here $V$ is the physical volume  of the fiducial cell
\be
V \, = \, \int \, \d^3 x \sqrt{\det q} \, = \, 4 \pi \, L_o r_o^2 |\bar p_b| |\bar p_c|^{1/2} = 4 \pi \, |p_b| |p_c|^{1/2} ~.
\ee
The eigenvalues $V_{\mu \tau}$ of $\hat V$ can be found  using (\ref{pb_pc_ev})
which yield $V_{\mu \tau} = 2 \pi \gamma^{3/2} \lp^3 |\mu| |\tau|^{1/2}$.

Classically, the field strength can be written in terms of holonomies using
\be\label{term2}
{}^o\!F_{ab}^k = - 2 \,\lim_{Ar_\Box
  \rightarrow 0} \,\, \Tr\,
\left(\f{h^{(\delta_{(i)}, \delta_{(j)})}_{\Box_{ij}}-1 }{\delta_{(i)} \delta_{(j)} \ell_{(i)} \ell_{(j)} } \right)
\,\, \tau^k\, \w^i_a\,\, \w^j_b\,  \ee
where
\be
h_{\Box_{ij}}^{(\delta_{(i)}, \delta_{(j)})} = h_i^{(\delta_{(i)})} h_j^{(\delta_{(j)})}(h_i^{(\delta_{(i)})})^{-1}(h_j^{(\delta_{(j)})})^{-1} ~.
\ee
In the quantum theory, due to underlying discreteness of quantum geometry, the loop $\Box_{ij}$ can only be shrunk to the minimum value of  area as given in LQG:
$\Delta = \beta \lp^2$ with $\beta$ of the order unity \cite{abl}. Hence the area of the loop
in $x-\theta$ and $x-\phi$ planes is constrained as:
\be\label{area1}
\delta_b r_o \, \delta_c  L_o = \Delta ~.
\ee
Unlike the loop in $x-\theta$ or $x-\phi$, the loop in $\theta-\phi$ is an open loop. However due to homogenity one can still associate an effective area with this loop \cite{aa:mb} and constrain it with $\Delta$:
\be\label{area2}
(\delta _b r_o)^2 = \Delta ~.
\ee
Using, Eqs. (\ref{area1}) and (\ref{area2}) we find
\be\label{dbdc}
\delta_b = \f{\sqrt{\Delta}}{r_o} ~, ~~ \mathrm{and} ~~ \delta_c =  \f{\sqrt{\Delta}}{L_o} ~.
\ee

Both $\delta_b$ and $\delta_c$  depend on length scales in the model. However,
the difference in their dependence is important.
The label of holonomies along equator and longitudes $\delta_b$, `knows' only about the radius $r_o$ of $\mathbb{S}^2$ where as $\delta_c$ which
labels holonomies along $x$  `knows' about $L_o$ -- length of the fiducial interval. This dependence is consistent with the expectation that holonomies along the equator and longitude should not feel the auxiliary length scale $L_o$.
As  will turn out, the relationship between $\delta_c$ and the auxiliary line interval $L_o$  plays an important role to obtain a consistent physics from this quantization.

{\it Remark:} Instead of constraining the area of the closed loops in $x-\theta$ and 
$x-\phi$ directions to be equal to $\Delta$, we could require them to be 
equal to the electric flux along the transverse edge, as in the case for a Bianchi-I model \cite{cs2,cs3}. For both of the loops 
it corresponds to equating $\db \dc L_o r_o$ with $2 \pi$ times the minimum eigenvalue of 
$\hat p_b$: $\db r_o \, \dc L_o = \pi \gamma \lp^2 \, \db$ which using 
(\ref{area2}) implies $\dc = \pi \gamma \lp^2/L_o r_o$. Though the numerical
factors change, the crucial dependence of $\delta_c$ on $L_o$ does not. In comparison to (\ref{area1}), in 
this case $\delta_c$ also feels the radius of $\mathbb{S}^2$.
In the present analysis, we will however work with Eqs.(\ref{area1}) and
 (\ref{area2}).

Combining terms (\ref{term1}) and (\ref{term2}) with computations of 
 the $\Omega$ term we get 
\ba\label{classicalC}
C^{(\db, \dc)} &=& \nonumber - \f{2}{\gamma^3 G \db^2 \dc} \Bigg[ 2 \, \gamma^2 \db^2 \mathrm{Tr} \left(\tau_3  h_x^{(\delta_c)} \, \{(h_x^{(\delta_c)})^{-1},V\}\right) \\
&&\hskip-1.25cm + \sum_{ijk} \varepsilon^{ijk} \mathrm{Tr}\left(h^{(\delta_{(i)}, \delta_{(j)})}_{\Box_{ij}}  h_k^{(\delta_{(k)})} \, \{(h_k^{(\delta_{(k)})})^{-1},V\} \right) \Bigg] ~.
\ea

We are interested in evaluating the corresponding symmetric operator: $\hat C_{\mathrm{SA}}^{(\db, \dc)} = (1/2) (\hat C^{(\db, \dc)} + 
\hat C^{(\db, \dc)\, \dagger})$, where $\hat C^{(\db, \dc)}$ obtained from (\ref{classicalC}) is
\begin{widetext}
\ba
\hat C^{(\db, \dc)} &=& \nonumber \f{32 \,i}{\gamma^3 \db^2 \dc} \Bigg[\sindb \cosdb \sindc \cosdc \left(\sindb \hat V \cosdb - \cosdb \hat V \sindb\right) \\
&&+\f{1}{2} \left(\sintdb \costdb + \f{1}{4}\gamma^2 \db^2\right)\left(\sindc \hat V \cosdc - \cosdc \hat V \sindc\right)\Bigg] ~.
\ea
\end{widetext}
We can now find the action of $\hat C_{\mathrm{SA}}^{(\db, \dc)}$ on the states $\Psi(\mu,\tau) = \langle\Psi|\mu,\tau\rangle$:
\begin{widetext}
\ba\label{qde}
\hat C_{\mathrm{SA}}^{(\db, \dc)} \Psi(\mu,\tau) &=& \nonumber \f{1}{2 \gamma^3 \db^2 \dc \lp^2} \Bigg[\left(V_{\mu + \db,\tau} - V_{\mu - \db, \tau} + 
V_{\mu + 3 \db,\tau + 2 \dc} - V_{\mu + \db, \tau + 2 \dc}\right)\, \Psi(\mu + 2 \db, \tau + 2 \dc) \\
&& \nonumber ~~~~~~~~~~~~~ + \left(V_{\mu - \db,\tau} - V_{\mu + \db, \tau} + 
V_{\mu + \db,\tau - 2 \dc} - V_{\mu + 3 \db, \tau - 2 \dc}\right)\Psi(\mu + 2 \db, \tau - 2 \dc) \\
&& \nonumber ~~~~~~~~~~~~~ + \left(V_{\mu - \db,\tau} - V_{\mu + \db, \tau} + 
V_{\mu - 3\db,\tau - 2 \dc} - V_{\mu - \db, \tau + 2 \dc}\right)\Psi(\mu - 2 \db, \tau + 2 \dc)\\
&& \nonumber ~~~~~~~~~~~~~ + \left(V_{\mu + \db,\tau} - V_{\mu - \db, \tau} + 
V_{\mu - \db,\tau - 2 \dc} - V_{\mu - 3 \db, \tau - 2 \dc}\right)\Psi(\mu - 2 \db, \tau - 2 \dc)\\
&& \nonumber ~~~~~~~~~~~~~ + \f{1}{2} \bigg[\left(V_{\mu,\tau + \dc} - V_{\mu, \tau - \dc} + 
V_{\mu + 4 \db,\tau + \dc} - V_{\mu + 4 \db, \tau -  \dc}\right)\Psi(\mu + 4 \db, \tau)\\
&& \nonumber ~~~~~~~~~~~~~~~~~~~ + \left(V_{\mu,\tau + \dc} - V_{\mu, \tau - \dc} + 
V_{\mu - 4 \db,\tau + \dc} - V_{\mu - 4 \db, \tau -  \dc}\right)\Psi(\mu - 4 \db, \tau)\bigg]\\
&& ~~~~~~~~~~~~~ + 2 (1 + 2 \gamma^2 \db^2) (V_{\mu,\tau-\dc} - V_{\mu,\tau+\dc}) \Psi(\mu,\tau)\Bigg] ~.
\ea
\end{widetext}
The states $\Psi(\mu,\tau)$ are further required to satisfy the invariance under parity operation: $\hat \Pi_b \Psi(\mu,\tau) = \Psi(-\mu,\tau)$. The quantum 
difference equation reduces to the one in of Ref. \cite{aa:mb} if we 
put $\db = \dc$ (except for the factor of $2$ multiplying   $(1 + 2 \gamma^2 \db^2)$ term). Thus various properties remain similar. These include its 
non-singular nature. If one considers 
$\tau$ as a clock, then the `evolution' occurs in the steps of $2 \dc$. By specifying the wavefunction at initial time steps $\tau = 2 n \dc$ and $\tau = 2(n - 1) \dc$, it is possible to backward evolve the equation through the 
classical singularity at $\tau = 0$. For that let us consider the case when 
$\mu = n \db$ where $n$ is positive integer greater than four\footnote{The character of the quantum difference equation changes for $n < 4$, however it 
still remains non-singular.}:
\ba
&& \nonumber (\sqrt{|\tau|} + \sqrt{|\tau + 2 \dc|}) \left(\Psi_{(n + 2) \db, \tau + 2 \dc} - 
\Psi_{(n - 2) \db, \tau + 2 \dc} \right) \\
&& \nonumber \hskip-0.2cm + \f{1}{2} \, (\sqrt{|\tau + \dc|} - \sqrt{|\tau -  \dc|}) \bigg[(n+2)\Psi_{(n + 4) \db, \tau} \\
&& \nonumber \hskip0.4cm ~~~~~ +  (n-2) 
\Psi_{(n - 4) \db, \tau}  - 2 n (1 + 2 \gamma^2 \db^2)  \Psi_{n \db,\tau} \bigg]\\
&& \nonumber \hskip-0.2cm + (\sqrt{|\tau|} + \sqrt{|\tau - 2 \dc|}) \left(\Psi_{(n - 2) \db, \tau - 2 \dc} - 
\Psi_{(n + 2) \db, \tau - 2 \dc} \right) \\
&& = 0 ~.
\ea
Since the coefficient of the term $(\Psi_{(n - 2) \db, \tau - 2 \dc} - 
\Psi_{(n + 2) \db, \tau - 2 \dc})$ never vanishes, we can use above equation to determine the wavefunction at negative values of $\tau$ starting the backward evoluution from the positive values. Thus we can `evolve' across the singularity.

It is to be emphasized that non-singular nature of above difference equation
should be only seen as a indication of the resolution of singularity in this
quantization. To have a detailed knowledge of the latter, it is important to
construct a physical Hilbert space and
finds expectation values of the Dirac observables (as accomplished for the 
isotropic LQC \cite{aps2}). This analysis will be performed elsewhere.

\section{Effective Dynamics}
\label{sec:5}

Let us analyse the dynamics resulting from
an effective Hamiltonian corresponding to the quantum constraint. This will be
on the lines of similar analysis in the isotropic LQC, where the effective 
Hamiltonian, derived using geometric methods in quantum mechanics, captures
the underlying quantum dynamics extremely well even for states which may not be sharply peaked \cite{nlqc2}. In the following, we assume that the 
effective spacetime description remains valid. 

It remains to be seen
if the same holds for the following effective Hamiltonian, which requires a detailed
numerical analysis  once we have the knowledge of 
physical states and observables in the model. The effective Hamiltonian is given by\footnote{In principle the effective Hamiltonian can have modifications coming from  the inverse powers of $|p_c|$ similar to in the case of inverse triads in the isotropic LQC. There its effects turn out to be negligible when compared with the corrections coming from field strength encoded in terms such as $\sin(\dc c)$. For this reason, we ignore these modifications in the following analysis.}
\ba\label{heff}
H_{\mathrm{eff}} &=& \nonumber - \f{N \mathrm{sgn}(p_c)}{2 G \gamma^2} \Bigg[2 \f{\sin (\delta_c c)}{\delta_c} \f{\sin (\delta_b b)}{\delta_b} \, |p_c|^{1/2} \\&& + \left(\f{\sin^2 (\delta_b b)}{\delta_b^2} + \gamma^2\right) \, p_b \, |p_c|^{-1/2} \Bigg] ~.
\ea
For $\db \rightarrow 0$ and $\dc \rightarrow 0$, $H_{\mathrm{eff}}$ approximates
$H_{\mathrm{class}}$ (\ref{h_cl}). 

To solve for the dynamics we choose 
$N = \gamma \mathrm{sgn} (p_c) |p_c|^{1/2} \delta_b/\sin(\db b)$ and
using (\ref{pbs}) we obtain \footnote{For analogous sets of effective dynamical equations arising in other prescription for Schwarzschild interior, see Refs. \cite{b-k,chiou,ksbound}.}  
\ba
\dot b &=& - \f{1}{2} \left(\f{\sin(\db b)}{\db} + \gamma^2 \f{\db}{\sin(\db b)}\right) ~,\\
\dot p_b &=& \f{1}{2} \, \cos(\db b) \left(1 - \gamma^2 \f{\db^2}{\sin^2(\db b)}\right) ~,\\
\dot c &=& - 2 \, \f{\sin(\dc c)}{\dc} \\
{\mathrm{and}} ~~~~ \dot p_c &=& 2 \, p_c \, \cos(\dc c) ~
\ea
where the derivatives are with respect to time $T$. Integrating the equations for $b$, $c$ and $p_c$, using the relations between $p_c^{(o)} = 4 m^2$, $l_o = 2m$ and $p_b^{(o)} = 2 m L_o$ and determining $p_b$ using $\H_{\mathrm{eff}} \approx 0$ we obtain
\be
b(T) = \pm \f{1}{\db} \, \cos^{-1}\bigg[b_o \tanh\left(\f{1}{2}\left(b_o T + 2 \tanh^{-1}(1/b_o)\right)\right)\bigg]
\ee
where
\be
b_o = (1 + \gamma^2 \db^2)^{1/2} ,
\ee
\ba
c(T) &=& \f{2}{\dc} \, \tan^{-1} \left(\mp \f{\gamma L_o \dc}{8 m} e^{-2 T}\right) ~,\\
p_c(T) &=& 4 m^2 \left(e^{2 T} + \f{\gamma^2 L_o^2 \dc^2}{64 m^2} e^{-2 T}\right)
\ea   
and 
\be
p_b = - 2 \f{\sin (\delta_c c)}{\delta_c} \f{\sin (\delta_b b)}{\delta_b} \f{|p_c|}{\f{\sin^2(\delta_b b)}{\delta_b^2} + \gamma^2} ~.
\ee
The effective dynamics predicts a non-singular bounce of $p_c$ and $p_b$ as the classical
singularity is approached. Of particular interest is the behavior of $p_c$. 
It takes a minimum value at time $T = (1/2) \ln(\gamma \Delta^{1/2}/8 m)$:
\be\label{pc_min}
p_{c \, \mathrm{(min)}} =   \gamma \, \Delta^{1/2} \, m
\ee
(where we have used Eq.(\ref{dbdc})). Since $\Delta = \beta\, \lp^2$, with $\beta$ is of order one,
we see that as $G \hbar \rightarrow 0$, $p_c \rightarrow 0$
\footnote{In the LQC literature $\beta$ is sometimes
taken as $\beta=4\sqrt{3}\,\pi\gamma$, but one can also see it as a free
parameter to be determined by physical considerations \cite{slqc}.}. 
Thus the 
resolution of Schwarzschild singularity has a pure quantum gravitational origin.

\begin{figure}[tbh!]
\includegraphics[width=8.5cm]{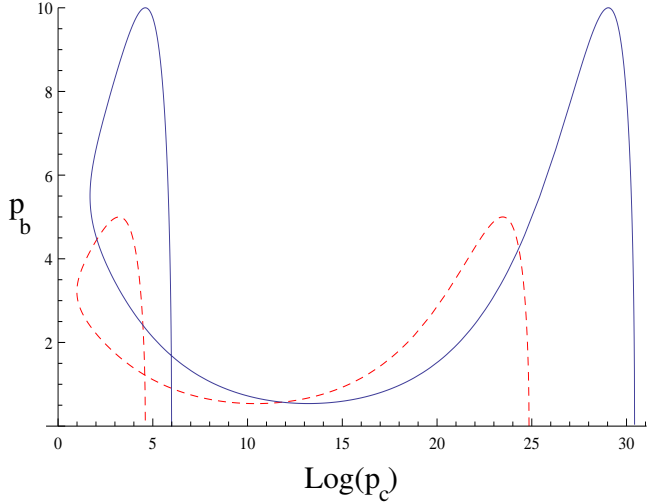}
\caption{Plot shows the evolution for different mass black holes with same $L_o$(equal to unity). Solid (black) curve shows the evolution for $m = 10$ and dashed (red) curve displays for $m=5$. Mass parameters are in  Planck units and we use the conventional values $\Delta = 4 \sqrt{3} \pi \gamma \lp^2$ and $\gamma = 0.2375$. } 
\label{fig:1}
\end{figure}

The variation of $p_b$ and $p_c$ obtained by solving the above effective equations is shown in figures 1 and 2. The evolution
starts from $p_b = 0$ at the horizon. Classically the evolution breaks down at the singularity 
where both $p_b \rightarrow 0$ and $p_c \rightarrow 0$. This does not happen
in the effective dynamics. 
In the high curvature regime, both $p_c$ and $p_b$ bounce 
and the singularity is avoided leading to a new white hole
spacetime. Fig. 1, shows the effects of changing the mass parameter on the evolution.  Fig. 2, displays the curves if one changes
the auxiliary length $L_o$. As can be seen, it only translates the 
curve vertically without changing the value of the mass of the white hole. At the low curvature scales,
the evolution approximates the classical theory.

To understand the details of the singularity resolution, let us consider the expansion ($\theta$) and shear ($\sigma^2$) scalars. These are given as follows:
\ba
\theta &=& \nonumber  \frac{1}{p_b} \frac{\d p_b}{\d \tau} + \frac{1}{2 p_c} \frac{\d p_c}{\d \tau} \\
&=& \nonumber \frac{1}{\gamma p_c^{1/2}} \Bigg[\frac{\sin(\delta_b b)}{\delta_b} (\cos(\delta_b b) + \cos(\delta_c c)) \\ && ~~~~~~~~~~~~~~ + 
\frac{\sin(\delta_c c)}{\delta_c} \cos(\delta_b b) \frac{p_c}{p_b} \Bigg]
\ea
and 
\ba
\sigma^2 &=& \nonumber \frac{1}{3} \left(\frac{1}{p_c} \frac{\d p_c}{\d \tau} - \frac{1}{p_b} \frac{\d p_b}{\d \tau} \right)^2 \\
&=& \nonumber \frac{1}{3 \gamma^2 p_c^{1/2}} \Bigg[\frac{\sin(\delta_b b)}{\delta_b} (2 \cos(\delta_c c) - \cos(\delta_b b)) \\ && ~~~~~~~~~~~  - \cos(\delta_b b) \frac{\sin(\delta_c c)}{\delta_c} \frac{p_c}{p_b} \Bigg]^2 ~.
\ea
In the classical theory, the expansion and shear scalars diverge at the central singularity where $p_b$ and $p_c$ vanish simultaneously. Since $p_c$ is bounded below, the
dynamical evolution in the effective spacetime description does not allow $\theta$ and $\sigma^2$ to diverge when the central singularity is approached in the loop quantized model. 
Thus, the expansion and shear scalars are dynamically bounded and the central singularity is resolved. The 
boundedness of expansion and shear scalars is an indication that geodesics are complete in this effective 
spacetime, past the would-be singularity, as was the case for the isotropic and anisotropic models in LQC 
\cite{ps09,ps11}. Finally, we note that at the horizon the expansion and shear scalars diverge both in the 
classical theory and the effective theory because of the coordinate singularity encountered there. 
\begin{figure}[tbh!]
\includegraphics[width=8.5cm]{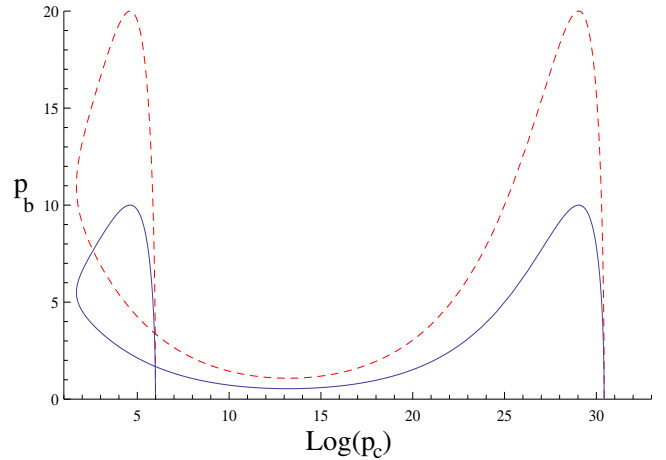}
\caption{Effects of changing $L_o$ are displayed in the figure. For both curves we choose $m=10$ in Planck units. 
The solid (black) curve corresponds to $L_o=1$ and the 
dashed (red) curve to $L_o=2$. The mass of the white hole formed after bounce does not depend on $L_o$ and is equal for both cases independent of the choice of parameters. 
} 
\label{fig:2}
\end{figure}

It is important to note that the expansion and the shear scalars are free from the underlying freedom of the rescaling of the fiducial length $L_o$. This is in contrast to the 
Ashtekar-Bojowald prescription, where these scalars turn out to be dependent on the fiducial length \cite{ksbound}. As has been stressed earlier by the authors, a consistency check of any quantization prescription is that the geometric scalars such as $\theta$ and $\sigma^2$ must be independent of any fiducial structure \cite{cs2,cs3}. Otherwise no 
consistent physical predictions on the details of the singularity resolution can be drawn from the quantization. So far, in the context of the black hole spacetimes the B\"ohmer-Vandersloot prescription was the only available quantization which satisfies this criteria \cite{ksbound}. Our quantization prescription is the only other choice which passes this criteria, {\it and} which leads to a mathematically and physically consistent dynamics. 
\begin{figure}[tbh!]
\includegraphics[width=8.5cm]{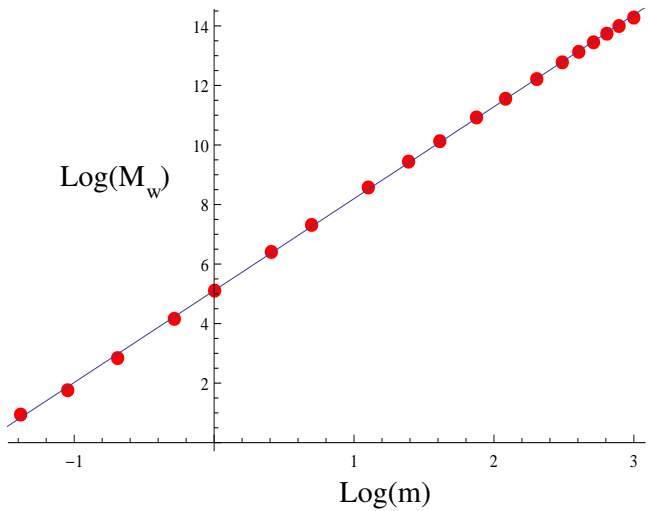}
\caption{The (log of the) mass of the white hole $(M_w)$ is plotted versus the (log of the)
initial black hole mass $m$. 
It turns out that the white hole mass is proportional to an approximately cubic power of the starting 
black hole mass. 
The fit to the data points obtained from effective dynamics is $\log(M_w) = 5.1086 + 3.08679 \log(m)$. }
\label{fig:3}
\end{figure}

An interesting aspect of the effective dynamics in this quantization prescription is that the mass $M_w$ of the white hole which forms after the singularity resolution is 
directly determined by the original black hole mass $m$. The relation, for different initial conditions, between the black hole mass and the resulting white hole mass 
is shown in Fig.~\ref{fig:3}. Phenomenologically, we find that the white hole mass is approximately proportional to the cube of the black hole mass. Furthermore, we can explicitly see from the
graph that the mass of the white hole is independent of the fiducial cell represented by $L_o$, thus
satisfying our criteria for consistency.

Let us compare the results with earlier works in more detail. 
For the quantization of the Schwarzschild interior 
in Ref. \cite{aa:mb}, an effective Hamiltonian was introduced in Ref. \cite{b-k} which can be written 
using (\ref{heff}) after replacing $\db$ and $\dc$ by a constant $\delta$. Note that departures from classical theory in the effective dynamics occur when trigonometric terms of the type $\sin(\delta c)$ in the effective Hamiltonian depart from 
$c$. 
The departures become most significant when the $\sin(\delta c)$ term saturates ($\sin(\delta c) \approx 1$). 
In the quantization proposed in Ref. \cite{aa:mb}, the saturation of  
$\sin(\delta c)$ is not independent of the choice of auxiliary structure ($L_o$) introduced to define the phase space, since for $L_o \rightarrow \alpha L_o$: $c \rightarrow \alpha c$ and $\delta$ does not scale.
(The term $\sin (\delta b)$ in the effective Hamiltonian though is independent of the rescaling in $L_o$). This is problematic as the dynamics then depends on the value of $L_o$ leading to unphysical effects.
 Hence, though we expect that fiducial structures should not affect physical predictions, the Ashtekar-Bojowald quantization prescription is an example where a theory fails this vital test. 

\begin{figure}[tbh!]
\includegraphics[width=8.5cm]{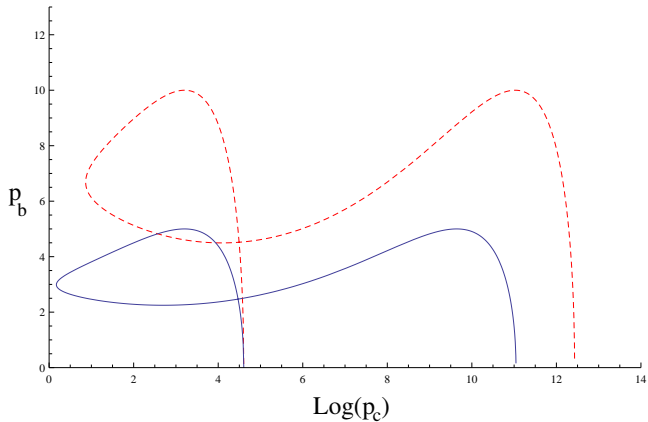}
\caption{This figure corresponds to the case for quantization of Schwarzschild interior in Ref. \cite{aa:mb}. Unlike the quantization presented in this manuscript (see Fig. 2), the mass of the white hole depends on the rescaling of $L_o$. 
For these simulations we have chosen $m = 5$ in Planck units. The solid (black) curve corresponds to $L_o=1$ and the 
dashed (red) curve to $L_o=2$. By changing the fiducial scale $L_o$, the mass of the white hole can be changed arbitrarily in the quantization presented in Ref. \cite{aa:mb}.
} 
\label{fig:4}
\end{figure}

Let us consider two instances in which the Ashtekar-Bojowald quantization yields inconsistent 
predictions. The first one is the minimum value of $p_c$ which
depends on $L_o$ \cite{b-k}. Since curvature scalars are proportional to inverse power of $p_c$, this implies that no sensible answer can be obtained for the 
upper bound on the value of spacetime curvature at which the bounce happens.
The second example is the relation between the black hole mass $m$ and the white hole mass $M_w$. It was noted earlier that in the Ashtekar-Bojowald prescription the white hole mass is governed by the fiducial length $L_o$ (through $p_b^{(o)}$) \cite{b-k}. 
Fig.~\ref{fig:4} explicitly demonstrates this difference between the two quantization prescriptions. We plot solutions obtained from the effective dynamics in the Ashtekar-Bojowald prescription for two different choices of $L_o$ for the same black hole mass. Since in this approach there is no distinction between $\delta_b$ and $\delta_c$ and they are assumed to be constant, we have chosen $\delta = 1$ in Fig.~\ref{fig:4}. We find that, in a striking comparison to 
the quantization put forward in this manuscript, the mass of the white hole changes with a change in $L_o$. Since the choice of the fiducial length $L_o$ is arbitrary, in the Ashtekar-Bojowald prescription the white hole can have an arbitrarily large or small mass which can be changed by a rescaling of $L_o$.

In contrast, in our approach the term $\sin(\dc c)$ in 
(\ref{heff}) does not depend on $L_o$. The reason is tied to the quantization
presented here, leading to Eqs.(\ref{dbdc}). These imply that under the change
$L_o \rightarrow \alpha L_o$: $\dc \rightarrow \alpha^{-1} \dc$, thus 
$\dc c \rightarrow \dc c$. Further, as in the previous case no effect is produced on the term 
$\sin(\db b)$ under the rescaling of $L_o$. The saturation of
$\sin (\dc c)$ and the resulting quantum gravitational effects are hence {\it independent} of the choice 
of $L_o$, which is seen from the plots in Fig. 2.
As a consequence of all this, the mass $M_w$ of the white hole in our analysis turns out to be 
independent of any fiducial structure. 

In summary, the reason for the success of our present quantization, when compared to the Ashtekar-Bojowald quantization, is the subtle difference in the way field strength is 
regulated in the quantum theory. The fact that subtleties in the quantization prescription can have
a deep impact on the physical predictions of the theory is not new. Indeed, we have seen something rather similar in loop quantum cosmology.
The old ($\mu_o$) quantization in LQC shared similar problems as the one in Ref. \cite{aa:mb}, i.e. unacceptable dependence of physics on auxiliary structures. It was 
cured by the introduction of the `improved' ($\bar \mu$) dynamics of LQC which resulted in
change of the uniform discreteness variable from triad to volume \cite{aps2}. In the 
Schwarzschild case, no such change is needed. The quantum difference equation
(\ref{qde}) is uniformly discrete in $p_c$ and $p_b$ as in the case of
the earlier quantization \cite{aa:mb}. Thus no refinement of the original lattice \cite{mb_cartin_khanna} 
or `improvement' on the lines of LQC \cite{b-k} are required. All such `improved' schemes
suffer from the problem of predicting ``quantum gravity'' effects at low curvatures near the coordinate 
singularity at the horizon. 
Such unphysical effects near the horizon are absent in the present quantization.


\section{Discussion}
\label{sec:6}

Let us summarize our results. In the process of quantizing the Schwarzschild interior,
we have noted that an important step in the Hamiltonian formulation was the understanding of the physical length scale $r_o$ from the very beginning. 
This scale corresponds, in the classical description,
precisely to the natural length scale in the system, the Schwarzschild radius. This simple
modification  has important ramifications. For, in the heart of the loop quantization, namely in the replacement of curvature by finite holonomies, there 
are two unequal parameters $\db$ and $\dc$. The former depending on $r_o$ and 
the latter on the auxiliary length scale $L_o$. As a consequence, the resulting Hamiltonian
operator is invariant under the change of fiducial cell. 
In this respect, the quantization found here can be seen as a refinement
of that in Ref.~\cite{aa:mb}. In particular, it reduces to that one when the parameter in each
direction are set to be equal\footnote{This would be however, a purely formal limit. For, as we have argued before, the dependence of these quantities on the two scales is distinct.}.
An important feature of the resulting quantum constraint is that it is perfectly regular at
the `would be singularity'. Physics does not stop there. 

Even when we do not have
a physical Hilbert space and Dirac operators thereon, one can expect that effective equations
describe the unitary evolution of appropriate semiclassical states, as happens in multiple examples within the isotropic sector. We 
have analysed in detail these effective equations and found interesting dynamics. In the relevant variable chosen the `initial 
condition' is given by the model universe (of spatial topology $\mathbb{S}^2\times\mathbb{R}$) approaching a null surface, that in 
the Schwarzschild solution corresponds to the (black hole) event horizon. The homogeneous hypersurfaces contract as they would
in Schwarzschild but, instead of reaching zero area in a finite proper time (the singularity),  the spheres reach a minimum value of 
area and `bounce back', approaching a different constant value in the asymptotic future. This asymptotic value (that could be called 
the mass of a white hole) only depends on the mass of the original black hole
and does not depend on the other degree of freedom available in the Hamiltonian description, nor on any fiducial structure. Analysis 
of the solutions of the effective equations show that the mass of the white hole is proportional to the cube of the starting black hole 
mass. In both asymptotic past and future, where the classical horizons might arise, the dynamics from the effective Hamiltonian 
approximates the classical dynamics. Thus, there are
no spurious quantum gravity effects appearing where one does not expect them. Our
quantization represents the first one possessing all these features. Further, analysis of the expansion and shear scalars shows that 
they are dynamically bounded on the approach to the classical singularity and are free from the rescaling of fiducial structures. 
These results stand in striking contrast to the resulting physics derived from the effective description of the Ashtekar-Bojowald 
quantization where the white hole mass, and the expansion and shear scalars depended on the fiducial length $L_o$. Finally,
we would like to end this note with two remarks.\\

\noindent{\it Relation with the Information loss issue.} The singularity resolution
found in Ref.~\cite{aa:mb} gave support to the paradigm proposed in \cite{aa-mb-info}
about black hole evaporation and information loss (further support comes from  work in the CGHS model 
\cite{cghs}). The basic idea is that the scenario in which a black hole looses
mass via Hawking radiation and evaporates is based in the assumption that there exists
a singularity inside the event horizon. If, as (loop) quantum gravity effects suggest, there is no
singularity, then one has to conclude that there was no event horizon to begin 
with and one has to extend 
the arena on which to describe the physical process, from the standard one with
a future singularity to some `quantum spacetime' containing no singularity. What can we say from our 
present study regarding this issue? The problem 
we have analysed is posed with the initial conditions corresponding to a space-like surface `just inside' 
the event horizon of the Schwarzschild spacetime, and through dynamical evolution, after the bounce
the asymptotic geometry to the future approaches that of a space-like surface `just inside' a
white-hole horizon of a different mass. 
While in the restricted sense of our analysis (vacuum and therefore 
no backreaction of matter), one obtains a different asymptotic region to the future, one can not 
conclude that the new paradigm does not hold. 
In order to settle this question, one would need a more complete treatment, including radiating matter, 
and an extension to the exterior region of the horizon. 
It is only with that more complete description that one might hope to 
give meaning to the `effective spacetime' that arises from the solution to our effective equations, 
in terms of the Ashtekar-Bojowald paradigm \cite{aa-mb-info}. 
In particular it should be clear that from the perspective of a complete non-singular quantum state,
statements  like `the mass of the initial black hole' or `final white hole' might become meaningless. 
Needless to say, a lot of work in this direction is needed, and
our present analysis can only be seen as a first step in that direction.\\

\noindent{\it Relation to $2+1$ Gravity.} In several instances, it has been useful to relate
3+1 symmetric spacetimes, via the Geroch reduction, to $2+1$ gravity coupled to a massless
scalar field $\psi$ \cite{geroch}. This equivalence has been exploited to study the quantum theory of Einstein Rosen waves \cite{ER} and polarized Gowdy models \cite{pierri}. For the
Kantowski-Sachs model under consideration here, this strategy is also a possibility that one might consider. To be precise, we have for the Schwarzschild interior all the ingredients needed in such reduction, namely, a Killing field $\xi=\partial/\partial x$ that is hypersurface orthogonal (to the $x=$constant surfaces). Thus, any
Kantowski-Sachs space-time (in vacuum) is equivalent to a spherically symmetric gravitational field $\tilde{g}_{ab}$ on a 3D spacetime with topology ${}^3\!{\cal M}=\mathbb{S}^2\times\mathbb{R}$, 
coupled to
a mass-less spherically symmetric scalar field $\psi$. The scalar field is proportional to the
logarithm of the norm of $\xi$ (given by $g_{xx}$). If we analyse the system from this perspective we see 
that the 3D Ricci curvature $\tilde{R}_{ab}$ (and energy density of the scalar field as defined by the 3D 
energy momentum tensor $\tilde{T}_{ab}$) diverge at both the  $t=0$ singularity and at the horizon where 
$t=2m$. The spurious curvature singularity at the horizon of the 3D metric $\tilde{g}$ disappears once we 
go back to the $3+1$ description by means of the conformal transformation that relates both metrics and 
that in this case `cures' the singularity,  yielding a finite 4D curvature at the horizon. It is to be 
noted that
any quantization that was build from the $2+1$ perspective  and that was successful in  
`curing' this 3D singularity would be, however, physically inadequate from the 4D perspective, where no 
classical singularity exists. In fact, what one might conclude is that the `improved' schemes of \cite{b-k} and \cite{cs2,cs3} are somewhat tailored to curing this 3D singularity as well, as can be seen from the 
fact that spurious quantum gravity effects appear near the horizon. In retrospect, this is not the first 
instance in which the classical equivalence between 4D symmetric models and $2+1$ gravity $+$ scalar field 
faces some difficulties in the quantum realm. In the case of polarized Gowdy models, the resulting 
quantization that used this equivalence in a fundamental way does not possess a unitary time evolution 
\cite{CCQ}. Furthermore, it has been shown that no such quantization exists for that field parametrization 
\cite{CMV}. In order to have a consistent, unitary, description one needs to take the 4D character of the 
problem at face value and find a suitable parametrization that does not admit a 
$2+1$ interpretation \cite{CCM}.

\section*{Acknowledgments}

\noindent 
We would like to thank A. Ashtekar for discussions and comments. 
This work was in part supported by DGAPA-UNAM IN103610 grant, by CONACyT 0177840 
and 0232902 grants, by the PASPA-DGAPA program, by NSF
PHY-1505411 and PHY-1403943 grants, and by the Eberly Research Funds of Penn State.

 \end{document}